# Generation of vortex-squeezed light in a coherently prepared medium

**FAN MENG, HAO ZHU, XIN-YAO HUANG†, GUO-FENG ZHANG***

*School of Physics, Beihang University, Beijing 102206, China*

*†Corresponding author:* xinyaohuang@buaa.edu.cn
**Corresponding author:* gf1978zhang@buaa.edu.cn

**In this Letter, we theoretically propose an alternative scheme for generating vortex optical squeezing, based on the Raman scattering process in a coherently prepared medium, distinct from approaches such as parametric down-conversion and four-wave mixing. Our analysis reveals that both the input control field and the generated signal field can exhibit squeezing upon adjusting the relevant system parameters, with the control field exhibiting a greater degree of squeezing under identical conditions. We further demonstrate the existence of optimal squeezing values over a range of tuning parameters, highlighting the flexibility and robustness of the proposed scheme. These findings may offer a new reference for continuous-variable vortex optical squeezing and possess potential applications in domains such as quantum information processing, quantum precision measurement, and quantum sensing.**

Quantum vacuum fluctuations of electromagnetic fields inherently produce quantum noise, the minimum level of which is constrained by Heisenberg's uncertainty principle and generally corresponds to the standard quantum limit (SQL) [1]. To surpass this limit, squeezed states have been extensively investigated as prototypical nonclassical optical fields. The squeezed state was first realized by Slusher et al. in 1985 via the four-wave mixing (FWM) process of Na atoms confined within an optical cavity [2]. In 1986, Shelby et al. generated squeezed light via FWM in optical fibers [3], and Wu et al. realized it through degenerate parametric down-conversion (PDC) in second-order nonlinear crystals within optical cavities [4]. Subsequently, extensive theoretical and experimental studies have been conducted on squeezed light [5-7]. The squeezed state remains the sole nonclassical state that can be prepared in a steady-state fashion, a property crucial for its deployment in measurement devices, as only the squeezed state enables continuous enhancement of measurement precision [5]. For instance, light squeezing can enhance the sensitivity of atomic magnetometry [8], optomechanical magnetometry [9], and laser interferometric gravitational-wave observatories [10].

Vortex light, a singular optical field capable of carrying orbital angular momentum (OAM), has garnered significant attention owing to its unique capacity to manipulate the Hilbert space dimension [11]. Squeezing of vortex light demonstrates superior performance in surpassing shot-noise limits and enhancing precision measurement accuracy [12]. Squeezing and entanglement of OAM can satisfy diverse application requirements and have found widespread use in areas such as quantum storage [13], quantum imaging [14], quantum metrology [15], and quantum communication [16]. Research on OAM states of discrete variables was first investigated [17,18], whereas studies of OAM states described by continuous variables have recently gained increasing attention owing to their advantages in deterministic preparation and high detection efficiency [19,20]. Currently, squeezed light is predominantly generated via PDC or FWM processes, necessitating phase-matching conditions and the use of strong optical fields to achieve sufficient optical nonlinearity [1,5]. Furthermore, studies on the squeezed states of vortex light typically first generate a fundamental-mode squeezed state, which is subsequently converted into an OAM squeezed state via a mode-conversion device, such as a spatial light modulator (SLM) [21]. This process may compromise both the conversion efficiency and the mode purity of the vortex light.

In this Letter, we theoretically propose an alternative scheme for generating vortex optical squeezing in a coherently prepared medium, based on the Raman scattering process and distinct from previous approaches such as PDC and FWM. Furthermore, we directly interact with the coherently prepared medium via vortex light, thereby avoiding the influence of mode conversion devices. We find that, by tuning the relevant system parameters, both the control and signal fields can be squeezed. This approach to generating continuous-variable vortex optical squeezing may offer a new reference and is anticipated to find applications in areas such as quantum information processing, quantum metrology, and quantum sensing.

As illustrated in Fig. 1, we consider an atomic ensemble with a three-level $\Lambda$ -type structure, in which the ground state is initially prepared in a coherent superposition. The control field $\Omega_c$ couples the energy levels $|g_2\rangle$ and $|e\rangle$, while the signal field $\Omega_s$ couples $|g_1\rangle$ and $|e\rangle$. The single-photon detunings are defined as $\Delta_c = \omega_c - \omega_{eg_2}$ and $\Delta_s = \omega_s - \omega_{eg_1}$, where ($\omega_c, \omega_s$) and ($\omega_{eg_1}, \omega_{eg_2}$) represent the corresponding light-field frequencies and the atomic transition frequencies, respectively. The two-photon detuning is given by $\delta = \Delta_s - \Delta_c$, a key parameter in the system. $\Gamma_1$ and $\Gamma_2$ denote the decay rates from the excited state to the respective lower states, and we assume $\Gamma_1 = \Gamma_2 = \Gamma/2$, where $\Gamma$ represents the total decay rate of the excited state $|e\rangle$. $\Gamma_{12}$ denotes the ground-state relaxation rate, representing both the population decay and the decoherence rate between ground states. In a suitable rotating frame, the system Hamiltonian can be expressed as:

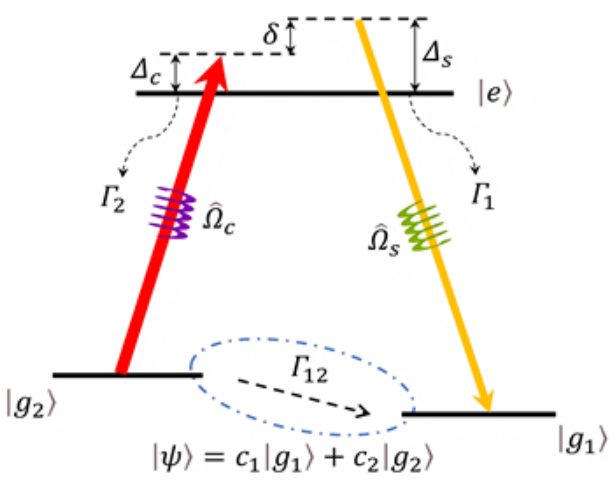

**Fig. 1.** Three-level $\Lambda$-type structure of a coherently prepared medium.

$$\hat{H} = -\hbar\Delta_s\hat{\sigma}_{ee}(z,t) - \hbar\delta\hat{\sigma}_{g_2g_2}(z,t) \\ -\frac{\hbar}{2}\left[\hat{\Omega}_s(z,t)\hat{\sigma}_{eg_1}(z,t) + \hat{\Omega}_c(z,t)\hat{\sigma}_{eg_2}(z,t) + \text{H.c}\right], \quad (1)$$

here, $\hat{\Omega}_s(z,t) = g_s\hat{E}_s(z,t)$ and $\hat{\Omega}_c(z,t) = g_c\hat{E}_c(z,t)$, where $g_s$ and $g_c$ denote the single-photon Rabi frequencies, with the assumption $g_s = g_c = g$; $\hat{E}_s$ and $\hat{E}_c$ are dimensionless operators representing the corresponding light fields.

The dynamic behavior of an atomic system is described by the Heisenberg-Langevin equation:

$$\frac{\partial}{\partial t}\hat{\sigma}_{\mu\nu} = \sum_j \Gamma_j\left(\hat{L}_j^\dagger\hat{\sigma}_{\mu\nu}\hat{L}_j - \frac{1}{2}\left\{\hat{\sigma}_{\mu\nu}, \hat{L}_j^\dagger\hat{L}_j\right\}\right) + \frac{1}{i\hbar}\left[\hat{\sigma}_{\mu\nu}, \hat{H}\right] + \hat{F}_{\mu\nu} \quad (\mu,\nu \in e, g_1, g_2), \quad (2)$$

where the first term on the right-hand side represents the dissipative dynamics, with $\hat{L}_j$ denoting the Lindblad operator associated with the $j$-th dissipation channel; the second term accounts for the primary dynamic evolution; and the third term corresponds to the Langevin noise operator.

The propagation of the light field within the medium is governed by the Maxwell-Schrödinger equations:

$$\left(\frac{\partial}{\partial z} + c^{-1}\frac{\partial}{\partial t}\right)\hat{\Omega}_s = i\frac{\Gamma\alpha}{2L}\hat{\sigma}_{g_1e}, \quad (3)$$

$$\left(\frac{\partial}{\partial z} + c^{-1}\frac{\partial}{\partial t}\right)\hat{\Omega}_c = i\frac{\Gamma\alpha}{2L}\hat{\sigma}_{g_2e}, \quad (4)$$

where $\alpha = 4g^2NL/c\Gamma$ denotes the optical density of the medium, $N$ represents the total number of atoms in the ensemble, and $L$ corresponds to the medium's length.

To solve the set of coupled Eqs. (1)-(4), we employ the mean-field approximation, decomposing the light field and atomic operators into their mean values and quantum fluctuation components. Consequently, the atomic operator can be decomposed as $\hat{\sigma}_{\mu\nu} = \sigma_{\mu\nu} + \hat{s}_{\mu\nu}$ $(\mu,\nu \in e, g_1, g_2)$, and the light field operator as $\hat{E}_\mu = E_\mu + \hat{a}_\mu$ $(\mu \in c, s)$, where $\sigma_{\mu\nu}$ and $E_\mu$ denote the mean-field components of the atom and the light field, respectively, while $\hat{s}_{\mu\nu}$ and $\hat{a}_\mu$ denote the corresponding quantum fluctuation operators. The quantum squeezing properties of the system are then analyzed through numerical solutions. The detailed solution procedure is provided in Supplement 1.

*Generation of squeezed vortex light*—the atomic ensemble is initially prepared in a coherent superposition of two ground states:

$$|\psi\rangle = c_1|g_1\rangle + c_2|g_2\rangle, \quad (5)$$

where we consider the weak interaction between the light field and the atomic ensemble [22], and, under this approximation, set $\sigma_{ee} \approx 0$, $\sigma_{g_1g_1} \approx |c_1|^2$, $\sigma_{g_2g_2} \approx |c_2|^2$, and $\sigma_{g_1g_2} \approx c_1^*c_2$. We investigate the Raman scattering process, assuming that initially only the control field $\Omega_c$ is present at the medium entrance $z = 0$, i.e., $\Omega_c(0) = \Omega$ and $\Omega_s(0) = 0$. Under these conditions, by solving the steady-state solution of the mean-field components of Eqs. (1)-(4), the propagation of the light field within the medium is determined as follows:

$$\Omega_c(z) = \frac{\Omega}{W}\left(\beta_2|c_2|^2 e^{-iWz} + \beta_1|c_1|^2\right), \quad (6)$$

$$\Omega_s(z) = \frac{\Omega}{W}c_1^*c_2\beta_1\left(e^{-iWz} - 1\right), \quad (7)$$

where $\beta_1 = \frac{\Gamma\alpha}{2L(2\Delta_s + i\Gamma)}$, $\beta_2 = \frac{\Gamma\alpha}{2L\left[2\Delta_c + i(\Gamma + \Gamma_{12})\right]}$ and $W = \beta_1|c_1|^2 + \beta_2|c_2|^2$.

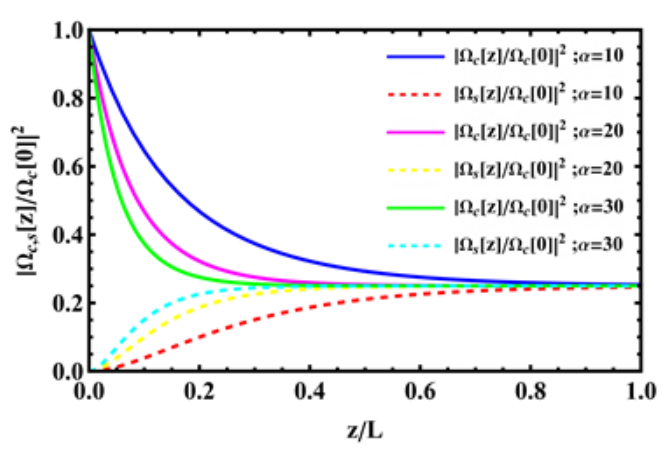

**Fig. 2.** Dependence of the normalized intensity of the control field $|\Omega_c(z)/\Omega_c(0)|^2$ and the generated signal field $|\Omega_s(z)/\Omega_c(0)|^2$ on the dimensionless length $z/L$. Other parameters are set as $\Gamma_{12} = 0$, $\delta = 0$, $c_1 = 1/\sqrt{2}$, $c_2 = 1/\sqrt{2}$, and optical density $\alpha = 10, 20, 30$.

If the initial control field $\Omega_c$ is a vortex beam, i.e., $\Omega_c(0) = \Omega = |\Omega|e^{il\varphi}$, where $l$ denotes the topological charge of the vortex light and $\varphi$ is the transverse azimuthal angle, then, considering the commonly used Laguerre-Gaussian vortex mode, the transverse profile of the vortex beam can be expressed as:

$$|\Omega| = \varepsilon\left(\frac{r}{w}\right)^{|l|} e^{-r^2/w^2}, \quad (8)$$

where $\varepsilon$ denotes the amplitude of the vortex beam, $r$ is the radial coordinate, and $w$ represents the beam waist. According to Eqs. (6) and (7), the generated signal field is given by $\Omega_s(z) \sim |\Omega|e^{il\varphi}$, indicating that, through the interaction between the control field $\Omega_c$ and the coherently prepared atomic ensemble, two vortex light beams with distinct frequencies can be generated at the medium's output. As illustrated in Fig. 2, the relationship between the normalized intensity of the control field $|\Omega_c(z)/\Omega_c(0)|^2$ and that of the generated signal field $|\Omega_s(z)/\Omega_c(0)|^2$ as a function of the dimensionless length $z/L$ is presented for optical densities $\alpha=10,20$ and $30$. It can be observed that higher optical densities allow the light field to establish electromagnetically induced transparency (EIT) conditions more rapidly within the medium; however, the optical density does not affect the final relative intensity of the light field $|\Omega_{c,s}(z)/\Omega_c(0)|^2$ once equilibrium (i.e., EIT) is reached.

To investigate the squeezing behavior of two vortex beams, we analyze the orthogonal variance $\Delta\hat{X}_\mu^2(\theta)$ $(\mu\in c,s)$ of the corresponding output fields, where $\Delta\hat{X}_\mu^2(\theta)=e^{-i\theta}\hat{a}_\mu+e^{i\theta}\hat{a}_\mu^+$ and $\theta$ denotes the orthogonal angle. By systematically scanning all orthogonal angles, an optimal orthogonal angle, $\theta_{opt}$, that maximizes the squeezing degree $V$ can be identified, with the squeezing degree defined by the following equation:

$$V\equiv\left\langle\Delta\hat{X}^2\left(\theta_{opt}\right)\right\rangle=-\left|\left\langle\hat{a}_\mu^2\right\rangle\right|-\left|\left\langle\hat{a}_\mu^{+2}\right\rangle\right|+2\left\langle\hat{a}_\mu^+\hat{a}_\mu\right\rangle+1\,. \quad \textbf{(9)}$$

By solving the quantum fluctuation components of Eqs. (1)-(4) and substituting them into the above expression, the quantum squeezing properties of the optical field can be investigated.

*Signal Field squeezing.* We first examine the squeezing associated with the generation of signal fields. Fig. 3(a) illustrates the variation of the generated signal field squeezing degree $V$ as a function of the two-photon detuning $\delta$ for optical densities $\alpha=10$, $20$, and $30$. For a given input control field $\Omega_c$ and optical density $\alpha$, an optimal two-photon detuning $\delta_{opt}$ exists that maximizes the signal field squeezing (i.e., minimizes $V$, denoted as $V_{opt,\delta}$), with higher optical densities leading to enhanced squeezing. To isolate the effect of two-photon detuning $\delta$ on squeezing, the ground-state relaxation rate is set to $\Gamma_{12}=0$. Fig. 3(b) depicts the variation of the signal field squeezability $V$ with the ground-state relaxation rate $\Gamma_{12}$ (setting $\delta=0$), for a given control field $\Omega_c$ and optical density $\alpha$. An optimal ground-state relaxation rate $\Gamma_{12,opt}$ exists that maximizes the signal field squeezing (i.e., minimizes $V$, denoted as $V_{opt,\Gamma_{12}}$), with higher optical densities enhancing the squeezing effect. This behavior arises because an intermediate $\Gamma_{12}$ facilitates signal field generation, whereas excessive $\Gamma_{12}$ leads to dissipation-dominated dynamics.

The above behavior can be theoretically explained. By analyzing the quantum fluctuation component, the fluctuation equation of the signal field $\Omega_s$ is derived as:

$$\frac{\partial}{\partial\xi}\hat{a}_s=P_1\hat{a}_s+Q_1\hat{a}_s^\dagger+R_1\hat{a}_c+S_1\hat{a}_c^\dagger+\hat{n}_s\,, \quad \textbf{(10)}$$

where $\xi=z/L$ denotes the dimensionless length, $P_1$, $Q_1$, $R_1$, and $S_1$ are the corresponding coefficients, and $\hat{n}_s$ represents the Langevin noise operator. $P_1$ and $Q_1$ correspond to the self-interaction processes of the signal field, whereas $R_1$ and $S_1$ correspond to the cross-interaction processes between the signal and control fields. These coefficients have distinct physical interpretations. $P_1$ denotes the self-phase damping process, which does not contribute to squeezing. $Q_1$ corresponds to the single-mode squeezing process, which is the primary mechanism for squeezing. Its contribution to the squeezing is given by $V=e^{-2|Q_1|\xi}$, assuming $Q_1$ remains constant with $\xi$. $R_1$ denotes the cross-phase coupling process, which likewise does not contribute to squeezing. $S_1$ corresponds to the dual-mode squeezing process, capable of generating entanglement. Finally, the presence of the Langevin noise operator $\hat{n}_s$ induces squeezing degradation, thereby reducing the overall squeezing. The optimal squeezing of the signal field arises from the competition between single-mode squeezing and other effects, including dual-mode squeezing, self-phase coupling, and cross-phase coupling. Although these interactions do not directly contribute to squeezing, they compete with the single-mode squeezing process, resulting in an optimal squeezing value $V$. Specifically, when both $\delta$ and $\Gamma_{12}$ are zero, standard EIT is established, rendering the medium fully transparent and the signal field unsqueezed. As $\delta$ and $\Gamma_{12}$ increase, the system deviates from standard EIT conditions, causing the medium to become partially opaque to the signal field and leading to quantum squeezing. However, when $\delta$ and $\Gamma_{12}$ become sufficiently large, dissipative processes dominate, weakening the squeezing and ultimately eliminating it.

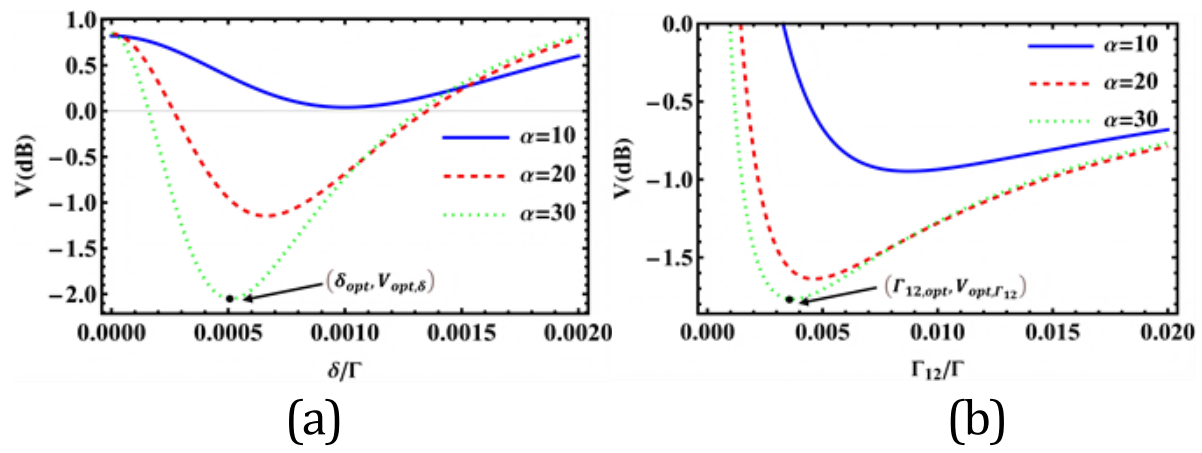

**Fig. 3.** (a) Dependence of the signal field squeezing degree $V$ on the two-photon detuning $\delta$ (with $\Gamma_{12}=0$). (b) Dependence of the signal field squeezing degree $V$ on the ground-state relaxation rate $\Gamma_{12}$ (with $\delta=0$). The remaining parameters in both panels are $\alpha=10,20,30$ and $\Omega_c=0.1\Gamma$.

*Control Field Squeezing.* Secondly, we examine the squeezing of the input control field. Fig. 4(a) illustrates the variation of the control field squeezing degree $V$ as a function of the two-photon detuning $\delta$ for optical densities $\alpha=10,20,30$ (with $\Gamma_{12}=0$). Concurrently, for a given input control field $\Omega_c$ and optical density $\alpha$, an optimal two-photon detuning $\delta_{opt}$ exists that maximizes the squeezing of the control field (i.e., minimizes $V$, denoted as $V_{opt,\delta}$), with higher optical densities enhancing the squeezing effect. Similarly, the optimal two-photon detuning $\delta_{opt}$ arises from the competition between single-mode squeezing and other system effects. Notably, the squeezing of the control field exceeds that of the signal field under identical conditions. This difference arises because, during the Raman scattering process, the signal field is passively generated via the coherently prepared medium, and its intensity $\Omega_s$ is lower than that of the incident control field $\Omega_c$. Fig. 4(b) depicts the variation of the control field squeezability $V$ with the ground-state relaxation rate $\Gamma_{12}$ (with $\delta = 0$) for a given control field $\Omega_c$ and optical density $\alpha$. In contrast to the signal field squeezing behavior in Fig. 3(b), the squeezing decreases rapidly with increasing $\Gamma_{12}$ and subsequently stabilizes. This behavior results from attenuation of the control field by the ground-state relaxation rate $\Gamma_{12}$ and the disruption of the coherent superposition.

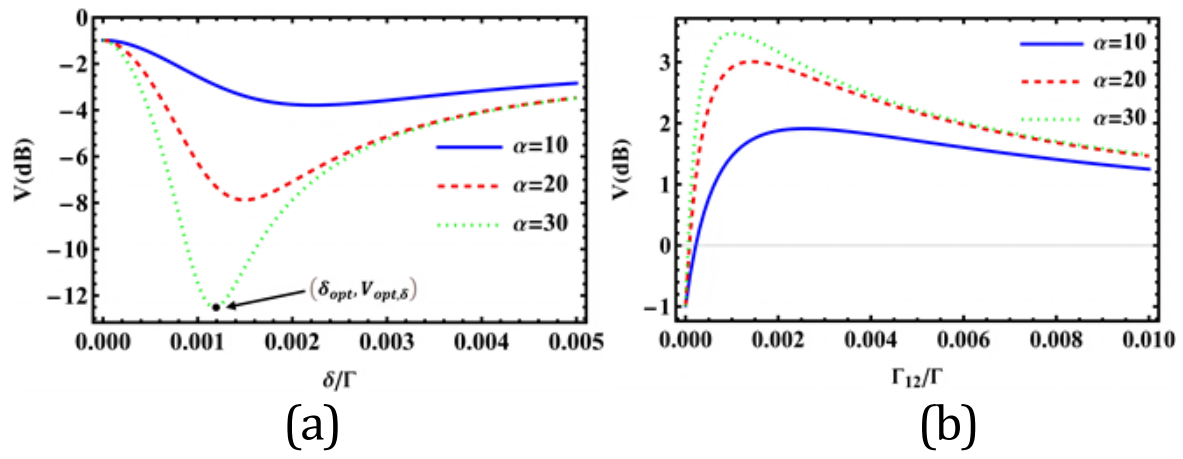

**Fig. 4.** (a) Dependence of the control field squeezing degree $V$ on the two-photon detuning $\delta$ (with $\Gamma_{12} = 0$). (b) Dependence of the control field squeezing degree $V$ on the ground-state relaxation rate $\Gamma_{12}$ (with $\delta = 0$). The other parameters are consistent with those in Fig. 3.

By analyzing the quantum fluctuation components of Eqs. (1)-(4), the fluctuation equation governing the control field $\Omega_c$ can be derived as:

$$\frac{\partial}{\partial \xi}\hat{a}_c = P_2\hat{a}_s + Q_2\hat{a}_s^\dagger + R_2\hat{a}_c + S_2\hat{a}_c^\dagger + \hat{n}_c . \quad \textbf{(11)}$$

Similarly, $P_2$ denotes the cross-phase coupling process, which does not contribute to squeezing; $Q_2$ corresponds to the dual-mode squeezing process, capable of generating entanglement; $R_2$ denotes the self-phase damping process, which does not contribute to squeezing; $S_2$ represents the single-mode squeezing process, which is the primary mechanism for squeezing. Its contribution is given by $V = e^{-2|S_2|\xi}$, assuming $S_2$ remains constant with $\xi$. The Langevin noise term $\hat{n}_c$ induces degradation of the squeezing. For the two-photon detuning, $|Q_1| \approx |S_2| \approx \alpha\Gamma\delta/(2\Omega^2)$; thus, when $\delta$ is small, the control field exhibits optimal squeezing. As $\delta$ increases, dissipative processes dominate, leading to the deterioration of squeezing. Regarding the ground-state relaxation rate, $|Q_1| \approx |S_2| \approx \alpha\Gamma\Gamma_{12}/(4\Omega^2)$; under these conditions, the squeezing trends of the control and signal fields differ significantly. This is because $\Gamma_{12}$ introduces a coupling term of order $\alpha\Gamma\Gamma_{12}/(4\Omega^2)$, whose dissipative component exceeds the squeezing contribution, resulting in predominant squeezing degradation in the control field. For the passively generated signal field, due to its low intensity $\Omega_s$, a smaller $\Gamma_{12}$ facilitates signal field squeezing, whereas an excessively large $\Gamma_{12}$ allows dissipative processes to dominate, leading to the existence of an optimal $\Gamma_{12,opt}$.

The coherent superposition of ground states proposed herein can be experimentally realized via fractional or partial stimulated Raman adiabatic passage (STIRAP) [23]. This three-level $\Lambda$-type configuration can be implemented in $^{87}Rb$ atoms. For instance, the excited state $|e\rangle$ may correspond to the $|5P_{1/2}, F=1, m_F=0\rangle$ state. The two ground states $|g_1\rangle$ and $|g_2\rangle$ may correspond to the $|5S_{1/2}, F=1, m_F=1\rangle$ and $|5S_{1/2}, F=1, m_F=-1\rangle$ states, respectively [24].

In conclusion, we theoretically introduce an alternative approach for generating vortex optical squeezing, based on the Raman scattering process in coherently prepared media, distinct from conventional methods such as PDC and FWM. We find that squeezing of both the control and generated signal fields can be realized by tuning relevant system parameters, with the control field exhibiting greater squeezing under identical conditions. The study further reveals that optimal squeezing values exist over a range of system tuning parameters, highlighting the flexibility and robustness of the proposed scheme. These findings may serve as a new reference for vortex optical squeezing and hold potential applications in areas including quantum information processing, quantum metrology, and quantum sensing.

**Funding.** This work was supported by the National Natural Science Foundation of China (Grant Nos. 12474353 and 12474354).

**Disclosures.** The authors declare no conflicts of interest.

**Data availability.** Data are available upon reasonable request.

**Supplemental Document.** See Supplement 1 for supporting content.